\documentclass[twocolumn,aps]{revtex4}
\usepackage{graphicx}

\begin{document}
\title{Bell's Inequality, Random Sequence, and
Quantum Key Distribution}

\author{Won-Young Hwang$^*$}

\affiliation{ Department of Physics Education, Chonnam National
University, Kwangjoo 500-757, Republic of Korea}

\begin{abstract}
The Ekert 91 quantum key distribution (QKD) protocol appears to be
secure whatever devices legitimate users adopt for the protocol, as
long as the devices give a result that violates Bell's inequality.
However, this is not the case if they ignore non-detection events
because Eve can make use of the detection-loophole, as Larrson
showed. We show that even when legitimate users take into account
non-detection events Eve can successfully eavesdrop if the QKD
system has been appropriately designed by the manufacturer. A
loophole utilized here is that of `free-choice' (or `real
randomness'). Local QKD devices with pseudo-random sequence
generator installed in them can apparently violate Bell's
inequality. \noindent{PACS: 03.65.Ud, 03.67.Dd}
\end{abstract}
\maketitle
Quantum key distribution (QKD) \cite{Ben84,Eke91,Gis02} is one of
the most promising protocols in quantum information processing
\cite{Nie00}. Besides the Bennett-Brassard 84 QKD protocol
\cite{Ben84,Gis02}, the interesting Ekert 91 (E91) protocol
\cite{Eke91,Gis02} makes use of the nonlocality of the
Einstein-Podolsky-Rosen pairs of quantum bits (qubits) \cite{Sel88}.

As is well known, even local realistic models can violate the Bell's
inequality \cite{Bel64,Sel88} if two remote experimenters, Alice and
Bob, simply do not take into account non-detection events in their
processing of the experimental data. This has been termed the
detection-loophole (e.g. \cite{Sel88,Ebe93,Hwa96} and references
therein. A discussion on the security of the Bennett-Brassard 84
protocol in connection with non-detection events has recently been
given \cite{Hwa04}.).

An issue in QKD is whether legitimate users (also Alice and Bob) can
trust a QKD system when they do not necessarily trust the
manufacturer of the system \cite{Lar02,May98}. Of course, this is
not a problem if Alice and Bob use only QKD systems that they have
made themselves. However, this is often impractical. If Alice and
Bob conclude that the QKD system provided by a manufacturer cannot
be trusted, the next problem is to determine how they can test the
QKD system as simply as possible. A basic ingredient for security of
the E91 protocol is the principle that an eavesdropper (Eve) cannot
emulate true entangled pairs of qubits by any local means, or more
specifically that Eve cannot violate Bell's inequality by any
separable states. One may think that this principle will hold
whatever detectors are used by Alice and Bob, including those
provided by the manufacturer, as long as the detectors give a result
that violates Bell's inequality. If this is the case, then the E91
protocol has an important advantage over other QKD protocols: Alice
and Bob do not have to test detectors carefully. However, the
principle can be violated apparently by a detection-loophole if
non-detection events are not taken into account, as said above.
Indeed, as shown by Larsson \cite{Lar02}, Eve can utilize the
detection-loophole in eavesdropping: A manufacturer who is a friend
of Eve designs a QKD system such that non-detection events are
ignored. If Alice and Bob use the system provided by the
manufacturer, then Eve can violate Bell's inequality apparently. In
this case, however, if Alice and Bob do not ignore non-detection
events then Eve can still be caught.

In this paper we strengthen the result of Ref. \cite{Lar02}. If the
manufacturer modifies the QKD system appropriately, Eve can
successfully eavesdrop even when Alice and Bob take into account
non-detection events. A loophole utilized here is that of
`free-choice' or `real randomness' (e.g.
\cite{Bel86,Kwi94,Hes01,Hes01-2,Gil02}).

This paper is organized as follows. We introduce the E91 protocol,
briefly describe Larsson's work, introduce the loophole of
free-choice and then show how the pseudo-random sequence can be
utilized by Eve to violate Bell's inequality by local means.
Finally, we briefly discuss a debate on Bell's inequality violation
\cite{Bel86,Kwi94,Hes01,Hes01-2,Gil02} and conclude.

In the E91 protocol, first Alice and Bob distribute $n$ pairs of
qubits in a Bell state $|\Psi^{-}\rangle= (1/\sqrt{2})(|0\rangle_A
|1\rangle_B - |1\rangle_A |0\rangle_B)$. Here $n$ is a positive
integer,  $|0\rangle$ and $|1\rangle$ are normalized and orthogonal
states, and $A$ and $B$ denote Alice and Bob. (The protocol here is
a modified version of the original E91 protocol with the essence
unchanged.) For each instance $i=1,2,...n$, each user randomly and
independently chooses to perform either normal measurement or
checking measurement. Later, the only cases used are those where
Alice and Bob's measurements match, while the other cases are
discarded. Among the matched cases, those where both Alice and Bob
choose a normal measurement is the normal phase, while the other
case is the checking phase. In normal the phase, each user performs
$S(z)$ measurements with a set of basis $\{|0\rangle,|1\rangle\}$.
The outcomes of $S_z$ measurements are perfectly correlated and thus
used as a key later. In the checking phase, each user performs those
measurements suitable for Bell's inequality violation of the state
$|\Psi^{-}\rangle$: Alice (Bob) randomly and independently chooses
one between the two directions $a$ and $a^{\prime}$ ($b$ and
$b^{\prime}$) for measurements. Then Alice (Bob) performs
spin-measurements in the chosen direction. Spin-measurement in
direction $p$ ($q$) is denoted as $S(p)$ ($S(q)$) where $p= a,
a^{\prime}$  ($q= b, b^{\prime}$). The probability that Alice gets a
result $\pm 1$ in spin-measurement $S(p)$ and Bob gets a result $\pm
1$ in spin-measurement $S(q)$ is denoted as $P_{\pm \pm}(p,q)$. The
correlation function $E(p,q)$ is given by $E(p,q)= P_{++}(p,q) +
P_{--}(p,q)- P_{+-}(p,q) - P_{-+}(p,q)$. Bell's inequality
\cite{Bel64,Sel88,Nie00} is then given by $|E(a,b)+ E(a,b^{\prime})+
E(a^{\prime},b)- E(a^{\prime},b^{\prime})| < 2$. Here the four
directions, $a, a^{\prime}, b,$ and $b^{\prime}$, are chosen such
that Bell's inequality is violated for the state $|\Psi^{-}\rangle$.

The idea of the E91 protocol can be summarized as follows. If Eve
provides separable states to Alice and Bob, then Eve can get
information on the key. However, in this case the separable states
cannot violate Bell's inequality in the checking phase. Thus Eve is
detected. For example, let us assume that Eve provides either
$|0\rangle |1\rangle$ or $|1\rangle |0\rangle$ with equal
probability after recording which state she sends at each instance.
In the normal phase nothing unexpected happens here. In the checking
phase, however, the samples cannot violate Bell's inequality and
thus the attack by Eve is detected. On the contrary, if Eve provides
the legitimate Bell state then Eve can pass the checking phase.
However, in this case Eve has no information on the key generated at
Alice's and Bob's sites. If Eve provides a partially entangled
state, then she will get partial information on the key.

Let us consider Larrson's point \cite{Lar02}. Classical information
can be encoded on the timing of pulses carrying qubits. The
manufacturer designs the QKD system such that the system can read
out and make use of the classical information thus encoded. Here
apparently the system works normally for Alice and Bob. Assume that
a hidden variable $\lambda$ is encoded in the classical information
part and that the system is programmed such that it violates Bell's
inequality by making use of the detection-loophole. Then Alice and
Bob will observe the system violating Bell's inequality in the
checking phase. Thus Eve is not detected.

However, we now describe another interesting case, termed 'the
free-choice loophole' \cite{Bel86,Kwi94,Hes01,Hes01-2,Gil02}: The
existence of real randomness is a necessary condition to derive
Bell's inequality. First, let us recall the locality condition for
Bell's inequality. Let $S_A(p)$ ($S_B(q)$) be an outcome of
spin-measurement in $p$ ($q$) direction at Alice's (Bob's) site.
Then a potential candidate is
\begin{equation}
\label{A} S_A(p)= f(p,q,\lambda), \hspace{2mm} S_B(q)=
g(p,q,\lambda),
\end{equation}
where $f$ and $g$ are normal functions. In Eq. (\ref{A}), the
possibility that measurement outcomes at one site depend on those at
the other site is not excluded. The reason why the model in Eq.
(\ref{A}) is usually excluded is that if two events are space-like
separated, then Alice's device has no way of obtaining information
on the choice of Bob's device on spin-measurement direction $y$ at
the instance when the spin-measurement is performed, and vice versa.
Therefore, the measurement outcomes of Alice (Bob) do not depend on
those of Bob (Alice). That is, Eq. (\ref{A}) reduces to
\begin{equation}
\label{B} S_A(p)= f(p,\lambda), \hspace{2mm} S_B(q)= g(q,\lambda),
\end{equation}
which is the locality condition. However, there is an important
tacit assumption in the reduction from Eq. (\ref{A}) to Eq.
(\ref{B}) \cite{Bel86,Kwi94,Hes01,Hes01-2,Gil02}: The choice of
Bob's device on spin-measurement direction $q$ is {\it random} so
that Alice's device cannot predict which one Bob's device will
choose at a certain instance, and vice versa. Otherwise, even if the
two measurement events are space-like separated, Alice's device can
calculate which direction Bob's device will choose, and vice versa.
Thus, effectively, Alice's (Bob's) device has information on which
direction Bob's (Alice's) device chooses at the instance when they
are performing the measurement. Hence in this case the reduction is
not valid in general.

Let us now see how Eve can utilize the free-choice loophole in the
E91 protocol: The manufacturer designs a QKD system such that each
device chooses spin-measurement directions according to a
pseudo-random sequence that is installed in the device beforehand.
Here the pseudo-random sequences in the two devices are independent.
The pseudo-random sequence is one that appears to be random but
actually is not. For example, the sequence 9869604401089... is
apparently random but it is obtained from $\pi^2$. The QKD system is
also designed such that one device contains an algorithm for
generating the pseudo-random sequence of the other device. Thus,
effectively, one device has information about the choices on
spin-measurement direction of the other device. Therefore, the
locality condition in Eq. (\ref{B}) can be effectively violated in
the QKD system provided by the manufacturer. In this case, users,
after careful inspections, will become aware of a problem in the
devices and that the measurement choices claimed to be random are
not really random, of course. However, it is impractical for many
users to perform such a careful inspection, and moreover it is a
very difficult task to identify a pseudo-random sequence. In other
respects, the design of the devices is in line with that of
Larrson's: In that the hidden variable $\lambda$ is encoded on the
timing of the pulses carrying qubits. The devices can read out and
make use of the classical information thus encoded.  Therefore, Eve
can successfully eavesdrop by adopting an effectively non-local
hidden variable model that simulates the Bell state
$|\Psi^{-}\rangle$.
Note that here we are dealing with an effectively non-local hidden
variable model that can simulate any entangled state.

Now, let us briefly discuss a debate on Bell's inequality violation
\cite{Bel86,Kwi94,Hes01,Hes01-2,Gil02}. It is clear that the above
QKD devices that are intrinsically local can show nonlocal behaviors
apparently and effectively. This kinds of
`local-but-apparently-nonlocal' models have already been discussed
by several authors, e.g. Bell \cite{Bel86} and Kwiat et al
\cite{Kwi94}. Recently, Hess and Philipp claimed to present a local
model that violates Bell's inequality \cite{Hes01,Hes01-2,Gil02}.
However, their model is in the same as the
`local-but-apparently-nonlocal' models. It is a fact that purely
local (deterministic) models without any randomness feeded outside
can violate Bell's inequality. However, a big problem is whether the
applicable scope of the `local-but-apparently-nonlocal' models can
be extended to macroscopic beings that can generate randomness, e.g.
humans. This rather philosophical question is beyond scope of this
paper.

In conclusion, at first glance the E91 protocol can be secure even
if Alice and Bob use any QKD system, as long as the system gives a
result that violates Bell's inequality. However, as shown by
Larrson, this is not the case if they ignore non-detection events
because Eve can use detection-loophole. We showed that Eve can
successfully eavesdrop, even when Alice and Bob take into account
non-detection events, if the manufacturer has designed the QKD
system appropriately. A loophole utilized here is that of
free-choice. We showed how a local QKD devices with a pseudo-random
sequence generator installed in them can apparently violate Bell's
inequality. We briefly discussed a debate on Bell's inequality
violation that is involved with a question on randomness.

I thank G. Weihs for helpful discussions.

\end{document}